\documentclass[unsortedaddress,twocolumn, nofootinbib,10pt]{revtex4}
\usepackage[utf8]{inputenc}
\usepackage{amsmath,empheq,color}
\usepackage{amsfonts}
\usepackage{amsthm}
\usepackage{amssymb}
\usepackage{graphicx}
\usepackage{hyperref}
\usepackage[normalem]{ulem}
\usepackage{epigraph}
\usepackage{mathrsfs}
\usepackage{bbm}
\usepackage{wrapfig}
\usepackage{lineno}
\usepackage{scalerel,stackengine}

\begin{document}

\title{On the covariance matrix for Gaussian states}

\author{Angel Garcia-Chung}
\email{alechung@xanum.uam.mx}

\affiliation{Departamento de F\'isica, Universidad Aut\'onoma Metropolitana - Iztapalapa\\
San Rafael Atlixco 186, Ciudad de M\'exico 09340, M\'exico}

\begin{abstract}
We show the explicit expression for the covariance matrix of general Gaussian states in terms of the symplectic group matrices. We discuss how the criteria to characterize squeezing and entanglement using the covariance matrix give rise to new criteria in the symplectic matrix elements used to construct the general Gaussian states.
\end{abstract}

\maketitle

\section{Introduction}

Entanglement is one of the main features of quantum mechanics, not only due its philosophical implications but also for its applications in many different areas \cite{bennett2000quantum, divincenzo1995quantum, bouwmeester1997experimental,  pan2001entanglement,  axline2018demand, adesso2007entanglement, horodecki2009quantum, richens2017entanglement}. For this reason, there are several criteria to determine whether a given state is entangled or not \cite{peres1996separability, horodecki1997separability, simon2000peres, duan2000inseparability, werner2001bound, giedke2001separability}. Among these criteria, those given by Simon \cite{simon2000peres}, Lu-Ming Duan et al \cite{duan2000inseparability} and Werner and Wolf \cite{werner2001bound} are particularly significant in the case of infinite-dimensional bipartite continuous variable  (CV) systems. 

A common aspect of these criteria is the use of the covariance matrix to detect entanglement, an idea first stated in \cite{simon2000peres}. There, Simon showed that separability forces a restriction on the covariance matrix that is stronger that the traditional uncertainty principle: if these restrictions on the covariance matrix hold then the state is entangled, and is not entangled otherwise. 

The states often used to derive the covariance matrix for CV systems are the Gaussian states, i.e., those states whose Wigner functions are a Gaussian functions on the phase space \cite{simon1988gaussian, braunstein2005quantum, walls2007quantum, adesso2014continuous}. A special class of Gaussian states is the class of Gaussian squeezed states which can be used to improve the sensitivity of measurement devices beyond the usual quantum noise limits \cite{walls1983squeezed, ma1990multimode, braunstein2005quantum, walls2007quantum, adesso2014continuous, schnabel2017squeezed}. These Gaussian squeezed states are generated by squeeze operators \cite{ma1990multimode} which can be considered as elements of the unitary representation of the symplectic group $Sp(2n, \mathbb{R})$, specifically, those close to the identity \cite{moshinsky1971linear, Arvind:1995ab, torre2005linear, wolf2016development}. 

The symplectic group is not just attached to the construction of the squeezed states. It is also manifest in Simon and in Werner-Wolf inseparability criteria. This leads us to the following question: is there a direct relation between the covariance matrices and the symplectic group elements? The answer to this question is in the affirmative and is the goal of this work.

We will show that the covariance matrices of Gaussian states are directly related with the symplectic group matrices. To do so, we briefly described in Section (\ref{preliminaries}) the mathematical tools concerning the symplectic group and its unitary representation. Then, in Section (\ref{CovarianceMatrix}) we calculate the explicit form of the covariance matrix in terms of the symplectic group elements. In Section (\ref{Discussion}) we discuss our results and provide some insights of its implications.

\section{Preliminaries} \label{preliminaries}
The symplectic group $Sp(2n, \mathbb{R})$ is formed by $2n \times 2n$ matrices ${\bf M}$ satisfying the equation
\begin{equation}
\left( \begin{array}{cc} {\bf 0} & {\bf 1} \\ - {\bf 1} & {\bf 0} \end{array}\right) = {\bf M} \left( \begin{array}{cc} {\bf 0} & {\bf 1} \\ - {\bf 1} & {\bf 0} \end{array}\right) {\bf M}^T, \label{SpCond}
\end{equation}
\noindent where ${\bf 1}$ is the $n\times n$ identity matrix and ${\bf M}^T$ is the transpose matrix of ${\bf M}$. Matrix ${\bf M}$ can be written in block form as follows
\begin{equation}
{\bf M} = \left( \begin{array}{cc} {\bf A} & {\bf B} \\ {\bf C} & {\bf D} \end{array}\right), \label{NoTildeMatrix}
\end{equation}
\noindent where ${\bf A}$, ${\bf B}$, ${\bf C}$ and ${\bf D}$ are $n \times n$ real matrices. The condition (\ref{SpCond}) now reads as
\begin{equation}
{\bf A} {\bf D}^T - {\bf B} {\bf C}^T = {\bf 1}, \quad {\bf A} {\bf B}^T = {\bf B} {\bf A}^T, \quad {\bf C}{ \bf D}^T = {\bf D} {\bf C}^T. \label{Sp1Coord}
\end{equation}

The symplectic group is used to provide the action of the linear canonical transformations on the phase space $(\mathbb{R}^{2n}, \{, \})$ where $\{, \}$ is the standard Poisson bracket. The group action is then given as
\begin{equation}
 \left( \begin{array}{cc} \vec{q}' & \vec{p}' \end{array}\right)^T = {\bf M}  \left( \begin{array}{cc} \vec{q} & \vec{p} \end{array}\right)^T, \label{1Coord}
\end{equation}
\noindent where $\vec{q}=(q_1, q_2, \dots, q_n)$ and $\vec{p}=(p_1, p_2, \dots, p_n)$ are the coordinates on the space $\mathbb{R}^{2n}$. Using these coordinates, the Poisson bracket can be written as
\begin{equation}
\{  \left( \begin{array}{cc} \vec{q} & \vec{p} \end{array}\right)^T , \left( \begin{array}{cc} \vec{q} & \vec{p} \end{array}\right) \} = \left( \begin{array}{cc} {\bf 0} & {\bf 1} \\ - {\bf 1} & {\bf 0} \end{array}\right). \label{PoissonB}
\end{equation}
\noindent Inserting (\ref{1Coord}) in the Poisson bracket definition (\ref{PoissonB}) yields (\ref{SpCond}). 

This way of defining the symplectic group action on the phase space is very useful to obtain a unitary representation on a Hilbert space. However, to study entanglement conditions a more convenient way on the phase space $\mathbb{R}^{2n}$ is required. Consider the array $\vec{R} = ( q_1, \, p_1 , \, q_2 , \, p_2 , \, \hdots , \, q_n , \, p_n )$, which is rather different to $\left( \begin{array}{cc} \vec{q} & \vec{p} \end{array}\right)$. Then similarly to (\ref{1Coord}) we consider the group action to be of the form
\begin{equation}
\vec{R}'^T = \widetilde{\bf M} \vec{R}^T, \label{2Coord}
\end{equation}
\noindent where $\widetilde{\bf M}$ is the new form of the symplectic matrix. We now insert expression (\ref{2Coord}), mutatis mutandis, in the Poisson bracket (\ref{PoissonB}) to obtain the following symplectic group condition
\begin{equation}
\left( \begin{array}{cccc} {\bf J} & {\bf 0} & \dots & {\bf 0} \\ {\bf 0} & {\bf J} & \dots & {\bf 0} \\ \vdots  & \vdots & \ddots & \vdots \\ {\bf 0} & {\bf 0} & \dots & {\bf J} \end{array} \right) = \widetilde{\bf M} \left( \begin{array}{cccc} {\bf J} & {\bf 0} & \dots & {\bf 0} \\ {\bf 0} & {\bf J} & \dots & {\bf 0} \\ \vdots  & \vdots & \ddots & \vdots \\ {\bf 0} & {\bf 0} & \dots & {\bf J} \end{array} \right)  \widetilde{\bf M} ^T, \label{2SpCond}
\end{equation}
\noindent where ${\bf J}$ is the $2\times 2$ matrix given by ${\bf J} = \left( \begin{array}{cc} 0 & 1 \\ -1 & 0 \end{array}\right)$. Matrices $\widetilde{\bf M}$ can also be written in block form as
\begin{equation}
\widetilde{\bf M} = \left( \begin{array}{cccc} {\bf A}_{11} & {\bf A}_{12} & \dots & {\bf A}_{1n} \\ {\bf A}_{21} & {\bf A}_{22} & \dots & {\bf A}_{2n} \\ \vdots & \vdots & \ddots & \vdots \\ {\bf A}_{n1} & {\bf A}_{n2} & \dots & {\bf A}_{nn} \end{array}\right) , \label{TildeMatrix}
\end{equation}
\noindent and the condition (\ref{2SpCond}) now reads on these block matrices as
\begin{equation}
{\bf J} = \sum^{j=n}_{j=1} {\bf A}_{i\, j} {\bf J} {\bf A}^T_{i\,j}, \quad  {\bf 0} = \sum^{k=n}_{k=1} {\bf A}_{i\, k} {\bf J} {\bf A}^T_{j\, k}, \label{Sp2Coord}
\end{equation}
\noindent for all $i \neq j$ in the second condition and $i, j =1,2,\dots , n$. Expressions (\ref{Sp1Coord}) and (\ref{Sp2Coord}) define different conditions for the symplectic matrices although, they describe the same Lie group \cite{adesso2014continuous}. Notice in (\ref{TildeMatrix}) that when the off-diagonal block matrices ${\bf A}_{i \neq j} = {\bf 0}$ the symplectic matrix becomes block-diagonal 
\begin{equation}
\widetilde{\bf M}= \mbox{diag}({\bf A}_{11}, {\bf A}_{22} , \dots, {\bf A}_{nn}), \label{Entang}
\end{equation}
\noindent and as a result of (\ref{Sp2Coord}) each matrix ${\bf A}_{ii} $ is an element of the symplectic group $ Sp(2,\mathbb{R})$.

Both group actions (\ref{NoTildeMatrix}) and (\ref{TildeMatrix}) are related via the transformation $\Gamma$ \cite{adesso2014continuous} as
\begin{equation}
{\bf M} = \Gamma \; \widetilde{\bf M} \; \Gamma^{-1}, \label{RelationbetweenMs}
\end{equation}
\noindent where $\Gamma$ is given by
\begin{equation}
 \left( \begin{array}{cc} \vec{q} & \vec{p} \end{array}\right)^T = \Gamma \vec{R}^T , \label{CTransformation}
\end{equation}
\noindent and is such that $\Gamma^T  = \Gamma^{-1}$.

The symplectic group is a non-compact group which requires an infinite Hilbert space for its unitary representation. Let us consider the Hilbert space ${\cal H} = L^2(\mathbb{R}^n, d\vec{x})$. The unitary representation of $Sp(2n,\mathbb{R})$ is the map ${\bf M} \mapsto \widehat{C}_{\bf M}$, where $\widehat{C}_{\bf M}$ is a unitary operator $\widehat{C}_{\bf M} \in {\cal L}(\cal H)$. Here, ${\cal L}(\cal H)$ is the space of linear operators over the Hilbert space ${\cal H}$. The map is given by \cite{moshinsky1971linear,  torre2005linear, wolf2016development}
\begin{equation}
\widehat{C}_{\bf M} \Psi(\vec{x}) = \int d\vec{x}' C_{\bf M}(\vec{x} , \vec{x}') \Psi(\vec{x}') =: \Psi_{\bf M}(\vec{x}), \label{UnitaryRep}
\end{equation}
\noindent for $\Psi(\vec{x}) \in {\cal H}$ and where the kernel of this integral operator is 
\begin{equation}
C_{\bf M}(\vec{x}, \vec{x}') = \frac{e^{ \frac{i}{2 \hbar} \left[ \vec{x}^T {\bf D} {\bf B}^{-1} \vec{x} - 2 \vec{x}'^T {\bf B}^{-1} \vec{x} + \vec{x}'^T {\bf B}^{-1} {\bf A} \vec{x}'\right]}}{\sqrt{ (2 \pi i \hbar)^n \det {\bf B}}} . \label{Kernel}
\end{equation}
\noindent The expression (\ref{UnitaryRep}) was also used to define $\Psi_{\bf M}(\vec{x})$ which is the state resulting from the group action on the state $\Psi(\vec{x})$.

Three comments about this representation: (i) it can be checked that $\widehat{C}_{\bf M}$ is a unitary operator and (ii) the factor $\det {\bf B}$ gives rise to a well define action if the matrix ${\bf B}$ is singular and (iii) it provides a unitary representation for the entire symplectic group and not just for those elements close to the group identity. 

\section{Covariance matrix} \label{CovarianceMatrix}

Let us proceed with the calculation of the covariance matrix. Consider the state 
\begin{equation}
| \Psi_{\bf M} \rangle = \widehat{C}_{\bf M} | 0 \rangle = \int d\vec{x} \; \Psi_{\bf M}(\vec{x}) \, | \vec{x} \rangle, \label{InitialS}
\end{equation}
\noindent where $| 0 \rangle$ is the state $|0\rangle = |0\rangle_1 \otimes | 0\rangle_2 \dots | 0 \rangle_n$.  The ket $| 0 \rangle_j$ is the vacuum state of the j-th quantum harmonic oscillator. These states are Gaussian states due to the kernel (\ref{Kernel}). 

To calculate the covariance matrix we first calculate the following amplitude
\begin{equation}
\langle \Psi_{\bf M} | \widehat{W}(\vec{a}, \vec{b}) | \Psi_{\bf M} \rangle = \langle \Psi_0 | \widehat{C}^\dagger_{\bf M} \widehat{W}(\vec{a}, \vec{b}) \widehat{C}_{\bf M} | \Psi_0 \rangle, \label{Amplitude}
\end{equation}
\noindent where $\widehat{W}(\vec{a}, \vec{b})$ is a Weyl-algebra generator whose representation on ${\cal H}$ is
\begin{equation}
\widehat{W}(\vec{a}, \vec{b}) \Psi(\vec{x}) = e^{\frac{i}{2 \hbar} \vec{a}^T \vec{b}} e^{\frac{i}{\hbar} \vec{a}^T \vec{x}} \Psi(\vec{x} + \vec{b}).
\end{equation}

The amplitude in  (\ref{Amplitude}) takes the following form
\begin{equation}
\langle \Psi_{\bf M} | \widehat{W}(\vec{a}, \vec{b})  | \Psi_{\bf M} \rangle  = \exp\left\{ - \frac{1}{4}   \left( \begin{array}{cc} \vec{a} & \vec{b} \end{array} \right)^T {\bf \Lambda} \left( \begin{array}{c} \vec{a} \\ \vec{b} \end{array}\right) \right\}, \label{Amplitude2}
\end{equation}
\noindent where the matrix ${\bf \Lambda}$ is given by
\begin{equation}
{\bf \Lambda} := {\bf M} \left( \begin{array}{cc} \frac{1}{\hbar^2} {\bf L}^2 & {\bf 0} \\ {\bf 0} & {\bf L}^{-2} \end{array}\right) {\bf M}^T, 
\end{equation}
\noindent and $ {\bf L} = \mbox{diag}(  l_1 , l_2 , \dots ,  l_n )$. Here, $l_j = \sqrt{\frac{\hbar}{m_j \, \omega_j}}$ are the characteristic lengths of the quantum harmonic oscillators.

The variance matrix ${\bf V}^{(2)}$, whose components can then be written as
\begin{equation}
{\bf V}^{(2)} = \left( \begin{array}{cc} \langle \Psi_{\bf M} | \widehat{x}_j \; \widehat{x}_k  | \Psi_{\bf M} \rangle & \langle \Psi_{\bf M} | \widehat{x}_j \; \widehat{p}_k  | \Psi_{\bf M} \rangle \\  \langle \Psi_{\bf M} | \widehat{p}_j \; \widehat{x}_k | \Psi_{\bf M} \rangle  & \langle \Psi_{\bf M} | \widehat{p}_j \; \widehat{p}_k   | \Psi_{\bf M} \rangle \end{array}\right), 
\end{equation}
\noindent can be obtained from (\ref{Amplitude2}) using the following relations
\begin{eqnarray}
\langle \Psi_{\bf M} | \widehat{x}_j \; \widehat{x}_k  | \Psi_{\bf M} \rangle = -\hbar^2 \partial^2_{ a_j a_k} \langle \Psi_{\bf M} | \widehat{W}(\vec{a}, \vec{b})  | \Psi_{\bf M} \rangle \vert_{\vec{a}, \vec{b}=0}, \label{EqCV1}\\
\langle \Psi_{\bf M} | \widehat{x}_j \; \widehat{p}_k  | \Psi_{\bf M} \rangle = -\hbar^2  \partial^2_{a_j b_k} \langle \Psi_{\bf M} | \widehat{W}(\vec{a}, \vec{b})  | \Psi_{\bf M} \rangle \vert_{\vec{a}, \vec{b}=0},\label{EqCV2}  \\
\langle \Psi_{\bf M} |  \widehat{p}_j \; \widehat{x}_k  | \Psi_{\bf M} \rangle = -\hbar^2  \partial^2_{b_j a_k} \langle \Psi_{\bf M} |  \widehat{W}(\vec{a}, \vec{b})  | \Psi_{\bf M} \rangle \vert_{\vec{a}, \vec{b}=0}, \label{EqCV3} \\
\langle \Psi_{\bf M} |  \widehat{p}_j \; \widehat{p}_k  | \Psi_{\bf M} \rangle = -\hbar^2 \partial^2_{b_j b_k} \langle \Psi_{\bf M} |  \widehat{W}(\vec{a}, \vec{b})  | \Psi_{\bf M} \rangle \vert_{\vec{a}, \vec{b}=0}. \label{EqCV4}
\end{eqnarray}

Notice that the matrix for the first order moments is zero due to the symmetry of the Gaussian function in (\ref{Amplitude2}). The resulting expression for ${\bf V}^{(2)} $ is
\begin{equation}
{\bf V}^{(2)}  = \frac{\hbar^2}{2}{\bf M} \left( \begin{array}{cc} \frac{{\bf L}^2}{\hbar^2} & {\bf 0} \\ {\bf 0} & {\bf L}^{-2} \end{array}\right) {\bf M}^T . \label{CVMatrix}
\end{equation}
In case we are working with quadratures, which are dimensionless operators, the matrix ${\bf \Lambda}$ takes the form
\begin{equation}
{\bf \Lambda}_q = {\bf M} {\bf M}^T,
\end{equation} 
\noindent and then, from (\ref{SpCond}) we have that 
\begin{equation}
{\bf M}^T = - \left( \begin{array}{cc} {\bf 0} & {\bf 1} \\ - {\bf 1} & {\bf 0} \end{array}\right) {\bf M}^{-1} \left( \begin{array}{cc} {\bf 0} & {\bf 1} \\ - {\bf 1} & {\bf 0} \end{array}\right),
\end{equation}
\noindent which can be used to check that matrix ${\bf \Lambda}_q$ is a symplectic matrix. The corresponding covariance matrix is given by
\begin{equation}
{\bf V}^{(2)}_q = \frac{1}{2} {\bf M} {\bf M}^T, \label{QCVMatrix}
\end{equation}
\noindent which is not a symplectic matrix due to the $1/2$ factor. This expression together with the expressions (\ref{Amplitude2}) and (\ref{CVMatrix}) constitute the main results of this paper. It is worth to recall that this result is valid for systems with n-degrees of freedom and Gaussian states of the form given in (\ref{InitialS}). 

\section{Discussion} \label{Discussion}

Due to most of the conclusions derived for ${\bf V}^{(2)}_q$ can also be applied to ${\bf V}^{(2)}$, let us focus our discussion on the covariance matrix ${\bf V}^{(2)}_q$.

(1) It can be immediately stated from (\ref{QCVMatrix}) that the covariance matrix ${\bf V}^{(2)}_q$ is positive definite, hence it is a {\it bona fide} covariance matrix. By construction this is an obvious conclusion however it serves to apply Simon's criterion. Additionally, Williamson's Theorem is also directly applied and the symplectic map diagonalizing ${\bf V}^{(2)}_q$ is $M^{-1}$. The resulting symplectic eigenvalues are $\kappa_j = 1/2$.

(2) The squeezing criteria can also be applied to (\ref{QCVMatrix}). In this case, consider the notation for ${\bf M}$ in which each row is written using spherical coordinates, for example, row j-th is of the form: $$ \left( \lambda_j \cos(\theta_{j 1}),  \lambda_j \sin(\theta_{j 1}) \cos(\theta_{j 2}) ,  \dots, \lambda_j \prod^n_i \sin(\theta_{j i}) \right).$$ Then the covariance matrix takes the diagonal form ${\bf V}^{(2)}_q = \frac{1}{2}\mbox{diag}(\lambda^2_1, \lambda^2_2, \dots, \lambda^2_{2n})$. Consequently, for any ${\bf M}$ whose components in spherical coordinates gives any $\lambda^2_j < 1/2$ then the corresponding unitary operator $\widehat{C}_{\bf M}$ is a squeeze operator. Moreover, an idea worth to be explored is whether every classical squeeze matrix can give rise to a quantum squeeze operator \cite{garcia2020squeeze}. 


(3) To check whether a given matrix ${\bf M}$ gives rise to an entangled state $| \Psi_{\bf M} \rangle$ we use the transformation $\Gamma$ and define the covariance matrix $\widetilde{\bf V}^{(2)}_q$ as follows
\begin{equation}
\widetilde{\bf V}^{(2)}_q = \Gamma^{-1}\,  {\bf V}^{(2)}_q \, \Gamma .
\end{equation}
\noindent Then, once we insert (\ref{QCVMatrix}) on this expression the covariance matrix takes the following form
\begin{equation}
\widetilde{\bf V}^{(2)}_q =  \frac{1}{2}  \widetilde{\bf M} \widetilde{\bf M}^T ,
\end{equation}
\noindent where $\widetilde{\bf M} $ is given by (\ref{RelationbetweenMs}). We now apply the Werner \& Wolf separability criterion to $\widetilde{\bf V}^{(2)}_q$ and obtain the following results:

 (i) if $\widetilde{\bf M} $ is a block-diagonal matrix like in (\ref{Entang}), then the state $| \Psi_{\bf M}\rangle$ is separable. As can be seen, this is a necessary condition but is not sufficient. 
 
  (ii) The necessary and sufficient condition is: the state $| \Psi_{\bf M}\rangle$ is separable iff the off-diagonal block terms of $\widetilde{\bf V}^{(2)}_q$ are null, which implies
\begin{equation}
\sum^n_{j=1} {\bf A}_{i j} {\bf A}^T_{k j} = {\bf 0}, \quad \forall i \neq k.
\end{equation} 

This result is remarkable as it links the classical canonical transformation $\widetilde{\bf M}$ with entanglement conditions which are strictly quantum features of the system. 

(4) Time-evolution of classical linear systems is given by symplectic matrices, say ${\bf H}(t)$. The unitary operator $\widehat{C}_{\bf H}$ provides the time-evolution of the system. Consequently, the time-evolution of a given state $| \Psi_{\bf M} \rangle $ is given as
\begin{equation}
| \Psi_{\bf M}(t) \rangle = \widehat{C}_{\bf H} | \Psi_{\bf M} \rangle = | \Psi_{\bf H \cdot M} \rangle.
\end{equation}
\noindent which yields the covariance matrix
\begin{equation}
{\bf V}^{(2)}_q(t) = \frac{1}{2} {\bf H}(t) \; {\bf M} \; {\bf M}^T \; {\bf H}^T(t).
\end{equation}
\noindent This relation allows us to directly compute the time-evolution of Gaussian states (\ref{InitialS}) as an algebraic calculation.

(5) It is already known that second order polynomial can be seen as elements of the Lie algebra of the symplectic  group. On the other hand, the squeeze operators used in the literature are mostly constructed using the exponential map of these operators but written in quadratures. This means that the squeeze operators are the unitary representation of some elements ${\bf M}$'s of the symplectic group, specifically, those elements in the subgroup close to the group identity. The result (2) shows that we can construct more general squeeze operators using group elements which are not close to the identity of the group. An example of this is the element of the form
\begin{equation}
{\bf M}_{nc} = \left( \begin{array}{cc} {\bf 0} & {\bf B}_{nc} \\ - {\bf B}^{-T}_{nc} & {\bf 0} \end{array}\right).
\end{equation}
\noindent where ${\bf B}_{nc}$ is an arbitrary $n\times n$ invertible matrix and $nc$ stands for ``not connected''. The covariance matrix ${\bf V}^{(2)}_{q (nc)}$ for this symplectic matrix is
\begin{equation}
{\bf V}^{(2)}_{q (nc)} = \frac{1}{2} \left( \begin{array}{cc} {\bf B}_{nc} {\bf B}^T_{nc} & {\bf 0} \\ {\bf 0} & ({\bf B}_{nc} {\bf B}^T_{nc} )^{-1} \end{array}\right) ,
\end{equation}
\noindent When written in spherical coordinates this gives a diagonal ${\bf V}^{(2)}_{q (nc)}$ which can shows squeezing. Moreover, it can also show entanglement according to result (3) even if ${\bf V}^{(2)}_{q (nc)} $ has a diagonal form on these coordinates for the group action.
 
(6) The relevance of the amplitude (\ref{Amplitude2}), is that it allows us to calculated the higher order variances, i.e., ${\bf V}^{(2n)}$. This can be done by applying the 2n-th derivatives analogous to the equations (\ref{EqCV1})-(\ref{EqCV4}). This will facilitate the analysis of the non-classicality of the states $| \Psi_{\bf M} \rangle$.

Finally, we want to add three more comments related with the implications of this result:

 (i) There are scenarios like in Loop Quantum Cosmology \cite{ashtekar2003mathematical, bojowald2010canonical, bojowald2011quantum} where the covariance matrix analysis is not possible. The reason for this is that the representation of the Weyl algebra is not weakly continuous. As a result, we need to consider other criteria to decide whether our state $| \Psi_{\bf M} \rangle $ is entangled or not. Therefore, we can use symplectic matrices satisfying the result (2) to construct squeezed states or the result (3) to construct entangled states on these scenarios \cite{Garcia-Chung:2020cag}. 
 
 (ii) We may think that all these results are anchored to the very construction of the state $| \Psi_{\bf M} \rangle$ in (\ref{InitialS}). If we consider a state given as $| \Phi_{\bf M} \rangle = \widehat{C}_{\bf M} | n_1, n_2, \dots, n_N \rangle $ the corresponding amplitude $\langle \Phi_{\bf M} | \widehat{W}(\vec{a}, \vec{b}) | \Phi_{\bf M} \rangle $ will be modified by a factor (related with the Hermite polynomials) but the exponential in (\ref{Amplitude2}) will be a global term for every $n_j$. Then the result (2) will be modified but the result (3), based on the Werner \& Wolf criterion, will hold.
 
 (iii) It is worth to consider whether a similar expression arises on discrete variable systems and also, how the manipulation of the matrix components can modified the amount of entanglement on both, discrete variable and CV systems.

\section{Acknowledgments}
I thank M. Berm\'udez-Monta\~na for her comments and helpful discussions.

\end{document}